\documentclass[conference]{IEEEtran} 

\usepackage{cite}
\usepackage{amsmath,amssymb,amsfonts}
\usepackage{algorithmic}
\usepackage{graphicx}
\usepackage{textcomp}
\usepackage{xcolor}
\usepackage{mathtools}
\def\BibTeX{{\rm B\kern-.05em{\sc i\kern-.025em b}\kern-.08em
    T\kern-.1667em\lower.7ex\hbox{E}\kern-.125emX}}

\usepackage{algorithm}
\usepackage{algorithmic}

\usepackage{color}
\usepackage{tablefootnote}

\usepackage{amsmath,amssymb,amsthm}
\usepackage{multirow}

\newcommand{\mbf}{\mathbf}

\usepackage{algorithm}
\usepackage{algorithmic}
\usepackage{longtable,tabularx}

\begin{document}

\title{Dynamical Update Maps for Particle Flow with Differential Algebra}

\author{

\IEEEauthorblockN{Simone Servadio,}
\IEEEauthorblockA{\textit{Iowa State University}, Ames, IA, USA}
\thanks{Dr. Simone Servadio, Assistant Professor, Department of Aerospace Engineering, servadio@iastate.edu}

}

\markboth{Transactions on Aerospace and Electronic Systems}%
{Shell \MakeLowercase{\textit{et al.}}: Bare Demo of IEEEtran.cls for IEEE Journals}

\maketitle
\thispagestyle{plain}
\pagestyle{plain}

\begin{abstract}
Particle Flow Filters estimate the ``a posteriori" probability density function (PDF) by moving an ensemble of particles according to the likelihood. Particles are propagated under the system dynamics until a measurement becomes available when each particle undergoes an additional stochastic differential equation in a pseudo-time that updates the distribution following a homotopy transformation. This flow of particles can be represented as a recursive update step of the filter. In this work, we leverage the Differential Algebra (DA) representation of the solution flow of dynamics to improve the computational burden of particle flow filters. Thanks to this approximation, both the prediction and the update differential equations are solved in the DA framework, creating two sets of polynomial maps: the first propagates particles forward in time while the second updates particles, achieving the flow. The final result is a new particle flow filter that rapidly propagates and updates PDFs using mathematics based on deviation vectors. Numerical applications show the benefits of the proposed technique, especially in reducing computational time, so that small systems such as CubeSats can run the filter for attitude determination.     
\end{abstract}


\IEEEpeerreviewmaketitle


\section{Introduction}
\IEEEPARstart{T}{he} estimation problem consists of optimally merging information from noisy observations with prior information, in terms of a distribution, to obtain a representation of the posterior probability density function (PDF). When the propagation of PDFs is included, we obtain a full filtering algorithm. 

In the linear and Gaussian case, the optimal solution is the Kalman filter \cite{kal}. However, most systems are nonlinear and require different techniques to obtain an approximation of the distributions: classic examples are the Extended Kalman Filter (EKF) \cite{gelb} and the Unscented Kalman Filter \cite{ut}.

Particles represent an accurate representation of the PDF, where their positions and weight approximate a continuous distribution as a probability mass distribution (PMF). When a measurement becomes available, particle filters modify the weights of the particles to match the posterior PDF, such as the Bootstrap Particle Filter (BPF) \cite{gordon1993novel} and the Gaussian Particle Filters (GPF) \cite{kotecha2003gaussian}. Sequential Importance Sampling  (SIS) tries to populate the posterior region with particles by introducing an importance distribution to sample from and successively modify their weight according to the likelihood \cite{servadio2024likelihood}.

Another approach of particle filters is to migrate particles from their initial (prior) location to the posterior, constituting particle flow filters \cite{daum2010exact,daum2016gromov}, and more recently \cite{dai2022design}. Particle flow filters use a homotopy transformation that modifies the prior to the posterior gradually. This gradual change can be interpreted as a recursive measurement update \cite{zanetti2011recursive}, which can be applied to every single particle of the ensemble \cite{michaelson2024particle}. 

This paper proposes a new type of particle flow based on the recursive update formulation derived in the Differential Algebra (DA) framework to gain computational advantages without decreasing accuracy. The DA approximates functions with their Taylor expansion series, and it is embedded with numerical integration and analytical differentiation \cite{damap}. Its use has been proven beneficial for space applications \cite{massari2018differential}, especially for uncertainty propagation \cite{valli2012gaussian} and quantification  \cite{servadio2021differential}. Indeed, DA has provided an approximation of the solution flow of a dynamical system, introduced as the state transition polynomial map (STPM) \cite{nonl}. The STPM has been exploited to propagate (and update) central moments \cite{ss, least}.

The new particle flow derived in this work exploits the DA representation of variables and the composition of the functions to propagate and update an ensemble of particles with one single mathematical operation: a computationally fast polynomial evaluation. The result is a fast particle flow filter that is not weighted down by the expansive numerical integration of particles. The resulting algorithm is applied to the attitude determination problem \cite{wertz2012spacecraft} for small spacecraft, where computational light filtering software is a necessity. 

The paper is organized as follows. A general explanation of particle flow filtering is offered in Section \ref{sec1}, with the derivation of the flow differential equations via recursive update derived in Section \ref{sec2}. Section \ref{sec3} presents the capabilities of DA, which are then applied to the recursive update in Section \ref{sec4}. Section \ref{sec5} shows how DA can also be employed in the prediction step of the filtering algorithm, with the possibility of combining polynomial maps, Section \ref{sec6}. Section \ref{sec7} shows the application to a toy problem, while Section \ref{sec8} applies the newly derived filter to the attitude determination problem. Lastly, conclusions are drawn in Section \ref{sec9}.

\section{The Particle Flow Update} \label{sec1}
Consider some prior knowledge of the state of a system, described according to the \textit{prior} distribution, $p_{\mbf x}(\mbf x)$, and a known nonlinear measurement model in the form of
\begin{equation}
    \mbf y = \mbf h(\mbf x) + \mbf v
\end{equation}
with Jacobian
\begin{equation}
    \mbf H = \dfrac{d \mbf h(\mbf x)}{d\mbf x}
\end{equation}
where $\mbf y$ is the $m$-dimensional measurement, $\mbf x$ is the $n$-dimensional state, and $\mbf v$ is zero-mean additive noise, assumed Gaussian with known covariance matrix $\mbf R$. Whenever a measurement outcome is provided, Bayes' rule provides the \textit{posterior} probability density function, which merges measurement information with the prior PDF according to
\begin{equation}
    p_{\mbf x}(\mbf x|\mbf y) = \dfrac{p_{\mbf x}(\mbf x) p_{\mbf y}(\mbf y|\mbf x)}{p_{\mbf y}(\mbf y)}
\end{equation}
where $p_{\mbf y}(\mbf y|\mbf x)$ is the conditional distribution of the measurement given the state, called \textit{likelihood}, and $p_{\mbf y}(\mbf y)$ is the normalizing constant, the marginal distribution of the measurement, such that the posterior integrates to unity in its domain. After taking the log function, the Bayes's formulation becomes
\begin{equation}
    \log p_{\mbf x}(\mbf x|\mbf y) = \log p_{\mbf x}(\mbf x) + \log p_{\mbf y}(\mbf y|\mbf x) - \log p_{\mbf y}(\mbf y)\label{log_Bayes}
\end{equation}

The particle flow measurement update provides the posterior distribution as an ensemble of particles that move iteratively. Starting from Eq. \eqref{log_Bayes}, particle flow defines a scalar homotopy parameter $0\leq \lambda \leq 1 $ and the relative log-homotopy in the form of
\begin{equation}
    \log p_{\mbf x}(\mbf x|\mbf y, \lambda) = \log p_{\mbf x}(\mbf x) + \lambda \log p_{\mbf y}(\mbf y|\mbf x) - \log K(\lambda) \label{log_bayes}
\end{equation}
where the normalization constant $K$ has been parametrized by $\lambda$ to ensure that the posterior distribution is a valid PDF that integrates to unity. It is evident that $p_{\mbf x}(\mbf x|\mbf y, 0)$ reduces to the prior distribution while $p_{\mbf x}(\mbf x|\mbf y, 1)$ is the posterior PDF, as it represents the classic Baye's formulation. Particles move starting from the null pseudo-time condition until $\lambda = 1$, where the prior has been transformed into the posterior. Thus, this update takes place in this pseudo-time domain, following a stochastic differential equation that defines the flow of the particles:
\begin{equation}
    d \mbf x = \mbf f (\mbf x,\lambda) d\lambda + \mbf B(\mbf x,\lambda)d\mbf w_{\lambda}
\end{equation}
where $\mbf f (\mbf x,\lambda)$ is the drift function, $\mbf B (\mbf x,\lambda)$ is the diffusion matrix function, and $d\mbf w_{\lambda}$ is a Weiner process that accounts for the stochastic nature of the differential equation. When $\mbf B (\mbf x,\lambda)$ is null, then the flow becomes deterministic, where each particle is governed by the following differential equation
\begin{equation}
    \dot {\mbf x}(\lambda) = \dfrac{d \mbf x}{d \lambda} = \mbf f (\mbf x,\lambda)
\end{equation}
Different approaches and techniques derived various formulations of the flow $\mbf f(\mbf x,\lambda)$, such as the Gromov flow \cite{daum2016gromov} and the exact flow \cite{daum2010exact}.

\section{Particle Flowing with Recursive Update} \label{sec2}
The implementation of particle flow via the homotopy transformation of the prior distribution into the posterior can be interpreted, from a different perspective, as a recursive measurement update with inflated noise. Following the information filter formulation, the information state is defined as 
\begin{align}
    \mbf S &= \mbf P^{-1} \\
    \mbf z &= \mbf S \mbf x
\end{align}
where $\mbf S$ is the information matrix, inverse of the state covariance matrix. The information update is \cite{assimakis2012information}
\begin{align}
    \hat{\mbf z}^+ &= \hat{\mbf z}^- + \mbf H^T \mbf R^{-1} \mbf y \\
    \mbf S^+ &= \mbf S^- + \mbf H^T\mbf R^{-1} \mbf H
\end{align}
which can be parametrized with the pseudo-time increments $\Delta \lambda$ to follow a recursive update formulation
\begin{align}
    \hat{\mbf z}_{i+1} &= \hat{\mbf z}_i + \Delta \lambda_i \mbf H^T \mbf R^{-1} \mbf y \\
    \mbf S_{i+1} &= \mbf S_i + \Delta \lambda_i\mbf H^T\mbf R^{-1} \mbf H
\end{align}
By moving the prior mean and covariance to the left hand side of the equation, it is possible to consider the infinitesimal change in those quantities: $\Delta \hat{\mbf z} =  \hat{\mbf z}_{i+1} -  \hat{\mbf z}_{i}$ and $\Delta \mbf S =  \mbf S_{i+1} -  \mbf S_{i}$ . As $\Delta \lambda \rightarrow d\lambda$ approaches zero, resulting in the limit definition of the derivative, we obtain a set of ordinary differential equations for the information state and covariance
\begin{align}
    \dot{\hat{\mbf z}} &= \dfrac{d \hat{\mbf z}}{d \lambda} = \mbf H^T \mbf R^{-1} \mbf y \label{zdot} \\
    \dot{\hat{\mbf S}} &= \dfrac{d \hat{\mbf S}}{d \lambda} = \mbf H^T\mbf R^{-1} \mbf H \label{Sdot}
\end{align}
However, we are interested in obtaining equations directly in the state space. To achieve so, consider the identity $\mbf I = \mbf P \mbf S$, where $\mbf I$ is the identity matrix, which derives to $\dot{\mbf I} = \mbf 0$. Then, according to the chain rule
\begin{equation}
    \dot{\mbf I} = \dot{\mbf P} \mbf S + \mbf P \dot{\mbf S} =  \mbf 0
\end{equation}
which leads to 
\begin{align}
    \dot{\mbf P} &= - \mbf P \dot{\mbf S} \mbf P \\
    &=  - \mbf P  \mbf H^T\mbf R^{-1} \mbf H \mbf P \label{Pdot}
\end{align}
after having substituted Eq. \eqref{Sdot}. Since $\mbf x = \mbf P \mbf z$, the state differential equation is recovered by substituting Eq. \eqref{Pdot} and Eq. \eqref{zdot}:
\begin{align}
    \dot {\mbf x} &= \dot {\mbf P} \mbf z + \mbf P \dot {\mbf z} \\
    &= - \mbf P  \mbf H^T\mbf R^{-1} \mbf H \mbf P \mbf S \mbf x +  \mbf P \mbf H^T \mbf R^{-1} \mbf y\\
    & = \mbf P \mbf H^T \mbf R^{-1} \left( \mbf y - \mbf H \mbf x \right)
\end{align}
Thus, we obtained a set of ODEs for the measurement update step of the filter in the pseudo-time $\lambda$, as derived in \cite{michaelson2024particle},
\begin{align}
    \dot {\mbf x}(\lambda) &= \mbf P \mbf H^T \mbf R^{-1} \left( \mbf y - \mbf H \mbf x \right)  \label{xdotfin}\\
    \dot{\mbf P}(\lambda) &=  - \mbf P  \mbf H^T\mbf R^{-1} \mbf H \mbf P \label{Pdotfin}
\end{align}
that require integration from 0 to 1. With this recursive update, each particle undergoes numerical integration and flows from its prior position to its posterior location.

\section{The Differential Algebra Approach}\label{sec3}
Differential Algebra (DA) is a differentiation technique that approximates nonlinear functions with their high-order Taylor expansion series. In this paper, the major use is the DA capability of expanding the dynamics in their Taylor series expansion around a well-selected center. The expansion order can be freely tuned according to the desired level of accuracy requested by the application, with the drawback of increasing the computational burden on the processing unit. This representation is carried out in a new computer environment: while functions are usually based on simple evaluation at specific points in the classical floating point (FP) array representation, they are represented as a matrix of coefficients and exponents in the DA framework. Thus, DA is another, more efficient, representation of state transition tensors (STT) \cite{stt}, where derivatives are stored as a series of monomials rather than hyper-dimensional matrices as in the STT.

A generic polynomial is initialized around a center, $\bar{\alpha}$, as
\begin{equation}
    \alpha(\delta\alpha ) = \bar{\alpha} + \delta\alpha 
\end{equation}
where $\alpha(\delta\alpha ) $ is the equivalent polynomial representation of the FP $\bar{\alpha}$, since $\alpha(0) = \bar{\alpha}$. In the DA framework, operations take place directly onto the polynomial, providing the transformed polynomial approximation of a nonlinear function. For example, consider a simple transformation such as
\begin{equation}
    \beta = \cos \alpha
\end{equation}
The FP equivalent of $\bar{\alpha}$ is $\bar{\beta} = \cos \bar{\alpha}$, while the first order DA representation is the polynomial 
\begin{equation}
    \beta(\delta\alpha ) = \cos \bar{\alpha} + \dfrac{d \cos  \alpha}{d  \alpha} \Bigg\vert_{\bar{\alpha}} \delta\alpha
\end{equation}
function of the variable $\delta\alpha$. It is evident that $\beta(0) = \bar{\beta}$. In order to evaluate the function at a separate location, $\bar{\alpha}_1$, the FP representation requires the full cosine evaluation, $\bar{\beta}_1 = \cos \bar{\alpha}_1$, while the DA first order approximation would compute the image as 
\begin{equation}
    \beta(\bar{\alpha}_1-\bar{\alpha}) = \cos \bar{\alpha} + \dfrac{d \cos  \alpha}{d  \alpha} \Bigg\vert_{\bar{\alpha}} (\bar{\alpha}_1-\bar{\alpha})
\end{equation}
which is a simple polynomial evaluation.

While trivial in such a simple example, this approach is particularly powerful for numerical integration. When used for numerical propagation of systems dynamics, many integrations can be replaced with faster polynomial evaluations. An ensemble of particles is propagated by integrating the DA representation of the state as a polynomial and subsequently evaluating the resulting polynomial at each particle's original deviation. This approximation is embedded, in this paper, into the dynamics propagation and into the particle flow. 

\section{The DA Recursive Particle Flow }  \label{sec4}
In the normal particle flow filtering, the flow ordinary differential equation is repeated for each particle, making the filter unfeasible and cumbersome for high dimensional systems with hard dynamics, as many particles are needed to adequately represent the state distribution. Therefore, the DA approach to solving ordinary differential equations comes in aid for those applications where computational time and efficiency matter. 

Given an ensemble of particles $\mbf x^{(i)}$ that describe the prior state distribution, with known mean $\hat{\mbf x}^-$ and covariance $\mbf P^-$, the particles' deviation from the mean of each sample is stored as 
\begin{equation}
    \delta \mbf x^{-(i)} = \mbf x^{(i)} - \hat{\mbf x}^- \quad  \forall \quad i = 1,\dots,N_p
\end{equation}
where $N_p$ is the total number of particles. The DA state polynomial is initialized as 
\begin{equation}
    \mbf x^-(\delta \mbf x ) = \hat{\mbf x}^- + \delta \mbf x
\end{equation}
and integrated alongside Eq. \eqref{Pdotfin} from 0 to 1, substituting Eq. \eqref{xdotfin}. The final result is the state polynomial at the flow completion 
\begin{equation}
    \mbf x^+(\delta \mbf x, \mbf y ) = \mathcal{M}_{0\rightarrow 1}^{\hat{\mbf x}^-}(\delta \mbf x, \mbf y )
\end{equation}
where $\mathcal{M}_{0\rightarrow 1}^{\hat{\mbf x}^-}$ indicates the polynomial flow map from $\lambda_{in} = 0$ to $\lambda_{fin} = 1$, centered at the prior mean $\hat{\mbf x}^-$ in the deviation variable $\delta \mbf x$ and function of the provided measurement $\mbf y$. This map is a complex polynomial, up to a selected arbitrary order, that expresses how deviations around the mean at $\lambda_{in}$ change to $\lambda_{fin}$.

The measurement update is completed by calculating each prior particle's new location according to their deviation vector, using polynomial evaluation as an approximation for the numerical integration
\begin{equation}
     \mbf x^{+(i)} = \mbf x^+(\delta \mbf x^{(i)} , \mbf y ) \quad  \forall \quad i = 1,\dots,N_p
\end{equation}
Therefore, $N_p$ numerical integrations have been substituted by the more efficient variable evaluation of a polynomial, whose accuracy is tuned by selecting the truncation order of the Taylor expansion series. The final estimate is provided as the mean of the particles. Having all the same importance weight:
\begin{align}
    \hat{\mbf x}^+ &= \dfrac{1}{N_p} \sum^{N_p}_{i = 1} \mbf x^{+(i)} \\
    \mbf P^+ &= \dfrac{1}{N_p} \sum^{N_p}_{i = 1} (\mbf x^{+(i)} - \hat{\mbf x}^+)(\mbf x^{+(i)} - \hat{\mbf x}^+)^T
\end{align}
are the mean and covariance representation of the state posterior distribution.

\section{The DA Ensemble Propagation}\label{sec5}
The particle flow has been substituted with the DA propagation of an ensemble of particles, a technique that is usually reserved from the prediction step of particle filters computed in the DA framework, such as in \cite{valli2013nonlinear} and \cite{servadio2021differential}. Indeed, the DA polynomial approach to solving ODEs originated in dynamic system applications. Consider the initial value problem 
\begin{equation} 
\label{sys}
    \begin{cases} 
        \dot{\mbf x}(t) = \mbf f(\mbf x(t), t) + \mbf w(t) \\
        \mbf x(0) = \hat{\mbf x}_0
    \end{cases}
\end{equation}
where $\mbf f$ is the nonlinear equation of motion of the system affected by random noise $\mbf w(t)$, with known initial condition $\hat{\mbf x}_0$ at time $t=0$. 

In the common particle flow filter, each particle would undergo integration until a measurement becomes available, ready to switch to the flow integration. Instead, the DA approach follows the polynomial approximation. Given the set of $N_p$ particles at time step $k$, $\delta \mbf x^{(i)}_k $, the deviation is stored prior to propagation
\begin{equation}
    \delta \mbf x^{(i)}_k = \mbf x^{(i)}_k - \hat{\mbf x}_k \quad \forall \quad i = 1,\dots,N_p
\end{equation}
where $\hat{\mbf x}_k$ is the current estimate. The DA state polynomial is initialized around the estimate 
\begin{equation}
    \mbf x_k(\delta \mbf x_k ) = \hat{\mbf x}_k + \delta \mbf x_k   \label{inipol}
\end{equation}
and it gets integrated to time step $k+1$ according to Eq. \eqref{sys}, 
\begin{equation}
    \mbf x_{k+1}(\delta \mbf x_k) = \mathcal{M}_{k\rightarrow k+1}^{\hat{\mbf x}_k}(\delta \mbf x_k)
\end{equation}
creating the so-called State Transition Polynomial Map (STPM) \cite{nonl} $\mathcal{M}_{k\rightarrow k+1}^{\hat{\mbf x}_k}(\delta \mbf x_k)$ centered at $\hat{\mbf x}_k$ in the $ \delta \mbf x_k$ variable, which represents the state deviation vector. Once again, the accuracy of the map is arbitrarily picked by selecting the truncation expansion orders of the Taylor polynomials. 

Analogously to the update, the propagated particles are evaluated via polynomial evaluation
\begin{equation}
     \mbf x^{(i)}_{k+1} = \mbf x_{k+1}(\delta \mbf x^{(i)}_k) \quad \forall \quad i = 1,\dots,N_p
\end{equation}
avoiding $N_p$ expensive numerical integrations. 

\section{Map Combination}\label{sec6}
Particle flow filters, like most filters, are divided into prediction and update steps. In the prediction step, the state PDF is propagated forward in time until sensors provide a measurement when the propagated prior is updated to the posterior during the pseudo-time $\lambda$.

The polynomial approximation of variables enables the possibility of combining functions and connecting the initial PDF at time step $k$ directly with the updated posterior at time step $k+1$. The final result is a single polynomial map, a composition of maps that integrates the state polynomial twice (propagation and flow) alongside the updated covariance. Starting from the initial state polynomial of Eq. \eqref{inipol}, the STPM and the polynomial flow map can be combined as
\begin{align}
    \mbf x^+(\delta \mbf x_k, \mbf y ) &= \mathcal{M}_{0\rightarrow 1}^{\hat{\mbf x}^-}(\mbf x_{k+1}(\delta \mbf x_k), \mbf y ) \\
    &= \mathcal{M}_{0\rightarrow 1}^{\hat{\mbf x}^-}(\mathcal{M}_{k\rightarrow k+1}^{\hat{\mbf x}_k}(\delta \mbf x_k), \mbf y )\\
    &= \mathcal{P}(\delta \mbf x_k, \mbf y )
\end{align}
which is a polynomial map that connects particles directly from the previous time step to the current one in their updated location at the end of the flow using, as input, the particles' original deviation. This map is adapted by the measurement outcome $\mbf y$, as it has embedded the influence of the likelihood distribution similar to the classic particle flow.

Therefore, the resulting complete filtering algorithm, the Differential Algebra Recursive Update Flow Filter (DARUFF), is reported on Algorithm \ref{algo1}. It shows the simplicity of the approach, where the creation of the map $\mathcal P$ is the only step that requires attention, as it includes state and covariance integration of the polynomials according to  Eq.  \eqref{xdotfin} and Eq.\eqref{Pdotfin}. Thanks to the single integration, the DARUFF scales extremely well when increasing the number of particles in the ensemble. 

\begin{algorithm}
\caption{Differential Algebra Recursive Update Flow Filter (DARUFF)}
\label{algo1}
\begin{algorithmic}[1]
\STATE \textbf{Initialization:} 
\STATE \hspace{0.5cm} $\delta \mbf x^{(i)}_0  \quad \text{  Initial Ensamble}$
\STATE \hspace{0.5cm} $\hat{\mbf x_0} \ \text{and} \ \mbf P_0$
\FOR{each time step $k$}
    \STATE \hspace{0.5cm} $\mbf x_k(\delta \mbf x_k ) = \hat{\mbf x}_k + \delta \mbf x_k $
    \STATE \hspace{0.5cm} $\delta \mbf x^{(i)}_k = \mbf x^{(i)}_k - \hat{\mbf x}_k$
    \STATE \hspace{0.5cm} $\mbf x^+(\delta \mbf x_k, \mbf y )=\mathcal{P}(\delta \mbf x_k, \mbf y )$
    \STATE \hspace{0.5cm} $\mbf x^{+(i)} = \mbf x^+(\delta \mbf x^{(i)} , \mbf y )$
    \STATE \hspace{0.5cm} $\hat{\mbf x}^+_{k+1} = \text{mean}(\mbf x^{+(i)}) $
    \STATE \hspace{0.5cm} ${\mbf P}^+_{k+1} = \text{cov}(\mbf x^{+(i)}) $
\ENDFOR
\end{algorithmic}
\end{algorithm}

\section{Range Measurement Example} \label{sec7}
Consider a Gaussian prior state estimate 
\begin{equation}
    \hat{\mbf x}^{-} = \begin{bmatrix}
        -3.5 & 0 
    \end{bmatrix}^T
\end{equation}
with error covariance 
\begin{equation}
    \mbf P^{-} = 
    \begin{bmatrix}
        1 & 0.5 \\ 0.5 & 1
    \end{bmatrix}
\end{equation}
A range measurement is given as
\begin{equation}
    y = || \mbf x|| + v
\end{equation}
where $v$ is the measurement noise, with distribution $v \sim \mathcal{N}(0,0.1^2)$. The numerical outcome received from the filter is $y = 1$, and it is desired to update the distribution, obtaining a representation of the true posterior using a particle flow approach. For comparison purposes, the ODE numerical integration and the DA integration of this measurement update flow update have been initialized with the same exact set of particles, consisting of 1000 particles drawn directly from the given prior PDF. 

\begin{figure}[!htb]
    \centering
    \includegraphics[width=1.0\linewidth]{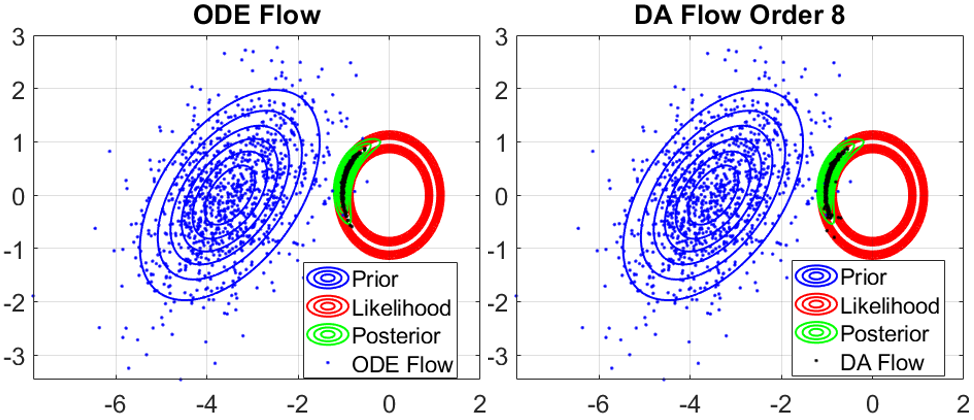}
    \caption{ODE vs. DA Flow solution}
    \label{fig:comp1}
\end{figure}
Figure \ref{fig:comp1} shows the prior, likelihood, and posterior distribution of the proposed application. The particles, starting at their given position (blue), flow to their final location to represent the posterior (black). The left figure is the ODE integration of the particles, while the right plot is its DA approximation, achieved with an 8th-order expansion. The two ensembles of particles behave very similarly, showing that the DA solution is a valid approximation of the PDF in terms of accuracy. 

\begin{figure}[!htb]
    \centering
    \includegraphics[width=1.0\linewidth]{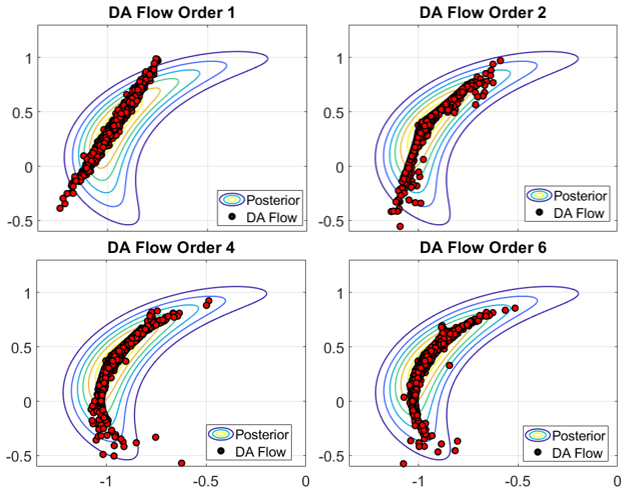}
    \caption{DA Flow Analysis with respect to the expansion order}
    \label{fig:order}
\end{figure}
The accuracy of the DA flow and the estimate depends on the selected arbitrary expansion order of the flow dynamics. Figure \ref{fig:order} shows how the ensemble of particles flows to its final position for different orders. As expended, increasing the order improves accuracy. The order 1 solution gives a linear distribution of particles, as it relies on mere linearization. Higher orders allow the ensemble shape to curve and turn, achieving a more accurate representation. It is worth remembering that the solved equations represent just the drift term of the flow dynamics, without diffusion. In the more general case, diffusion can be added as in \cite{michaelson2024particle}, spreading the particles accordingly (and randomly) to cover the thickness of the posterior PDF.

\begin{figure}[!htb]
    \centering
    \includegraphics[width=1.0\linewidth]{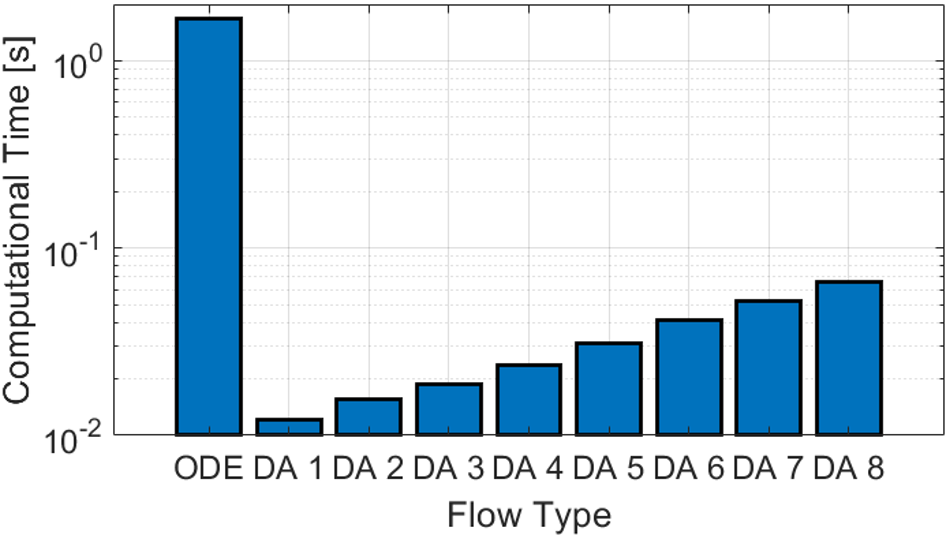}
    \caption{Computational Time Comparison}
    \label{fig:time}
\end{figure}
The main advantage of using DA over numerical integration is shown in Fig. \ref{fig:time}, where the requested computational time to propagate the flow using a Runge-Kutta 7/8 integrator is reported in a bar graph. The figure, in logarithmic scale, shows how much faster the DA integration is when compared to the regular approach. The DA flow computational time increases as the order becomes larger, but it is more than one order of magnitude (two for $c=1$) faster. Thus, the DA approximation with polynomial evaluations has a similar accuracy level but is drastically faster than selecting numerical propagation.

\section{Attitude Determination} \label{sec8}
Consider the following attitude determination problem. The quaternions $\mbf q$ of a CubeSat behave according to the following equations of motion
\begin{equation}
    \dot {\mbf q} = \dfrac{1}{2} \begin{bmatrix}
        \boldsymbol{\omega}_b \\ 0
    \end{bmatrix} \otimes
    \mbf q
\end{equation}
where $\boldsymbol{\omega}_b$ is the satellite angular velocity in body coordinates and $\otimes$ indicates the quaternions multiplication. The angular velocity follows the Euler equations
\begin{equation}
    \dot {\boldsymbol{\omega} }_b = \mbf J^{-1}\big( \mbf m - \boldsymbol{\omega}_b \times  \mbf J \boldsymbol{\omega}_b\big)
\end{equation}
where $\mbf m$ represents the vector of external torques and $\mbf J$ is the inertia matrix of the satellite, defined as $\mbf J = \text{diag}(\begin{bmatrix}100 &60 &50\end{bmatrix}) \text{kg} \text{m}^2$  in the current example. 

The satellite is equipped with a gyroscope and two star trackers, that provide measurements with a low frequency of 0.5 Hz. The star trackers point at two different stars, with inertial vectors $\mbf r_1$ and $\mbf r_2$ defined as 
\begin{align}
    \mbf r_1 &= \begin{bmatrix}5 &2 &3\end{bmatrix}^T \\
    \mbf r_2 &= \begin{bmatrix}1 &10 &4\end{bmatrix}^T
\end{align}
normalized such that they are unit vectors. The resulting measurement model is a quadratic function of the quaternions, as each star tracker vector is evaluated as 
\begin{align}
    &\quad \quad \quad \quad \quad  \mbf y_j =  \mbf C(\mbf q) \mbf r_j + \mbf v_j \quad\text{for} \ j = 1,2 \\
    &\quad \quad \quad \quad \mbf C(\mbf q) = \nonumber \\
    &\begin{bmatrix}
        q_s^2 + q_i^2 - q_j^2 - q_k^2 & 2(q_i q_j + q_s q_k) & 2(q_i q_k - q_s q_j) \\
2(q_i q_j - q_s q_k) & q_s^2 - q_i^2 + q_j^2 - q_k^2 & 2(q_j q_k + q_s q_i) \\
2(q_i q_k + q_s q_j) & 2(q_j q_k - q_s q_i) & q_s^2 - q_i^2 - q_j^2 + q_k^2
    \end{bmatrix}
\end{align}
where $q_s$ is the scalar component of the quaternion and $q_{i,j,k}$ the vectorial entries. Matrix $ \mbf C(\mbf q)$ is the Direct Cosine Matrix (DCM) and $\mbf v_j$ is zero-mean Gaussian noise with covariance $0.01^2\mbf  I_3$. The gyroscope directly measures the satellite's angular velocity as
\begin{equation}
    \mbf y_3 = \boldsymbol{\omega}_b + \mbf b + \mbf v_3
\end{equation}
where $\mbf b$ indicates bias in the measurement, while $\mbf v_3$ is zero-mean Gaussian noise with covariance $(0.2\pi/180)^2\mbf  I_3 \ \ \text{(rad/s)}^2$.

Therefore, the resulting state vector has dimension $n=10$, defined as
\begin{equation}
    \mbf x = \begin{bmatrix}
        \mbf q^T & \boldsymbol{\omega}_b^T & \mbf b^T
    \end{bmatrix}^T
\end{equation}
as it is desired to estimate the gyro's bias, assumed constant during the whole duration of the simulation. The selected initial condition is:
\begin{align}
    \mbf q_0 &= 0.5\ \begin{bmatrix}1 &1 &1 &1\end{bmatrix}^T\\
    \boldsymbol{\omega}_{b0} &= (10\pi/180)\dfrac{\begin{bmatrix}1 &2 &3 \end{bmatrix}^T}{||\begin{bmatrix}1 &2 &3 \end{bmatrix}^T||} \\
    \mbf b_0 &= \begin{bmatrix}0 &0 &0 \end{bmatrix}^T
\end{align}
The simulation is run for a total of 2 minutes and $250n$ total particles in the ensemble. The particle filter utilizes a fixed step 4th-order Runge-Kutta (RK4) integrator to propagate the particles and their flow. This selection is driven by ensuring a fair comparison in the computational time analysis between methodologies and in the number of function evaluations requested by the algorithm. Therefore, the RK4 is integrated with 0.01 seconds constant time step, while the flow integration in the pseudo-time follows a geometric distribution from 0.001 to 1 divided into 50 steps. The geometric distribution guarantees a correct initial movement of the particles at the beginning of the flow, with small steps, and larger propagation towards $\lambda =1$, where the flow is smaller, thus avoiding pointless calculations. 

\begin{figure}[!htb]
    \centering
    \includegraphics[width=1.0\linewidth]{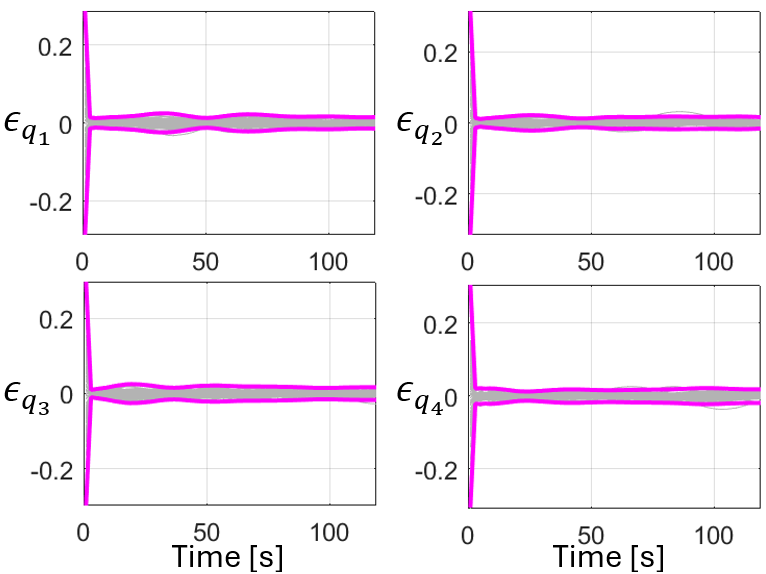}
    \caption{Monte Carlo analysis for the quaternions error.}
    \label{fig:monteq}
\end{figure}
The DARUFF attitude determination results are reported in Fig. \ref{fig:monteq}, where a Monte Carlo analysis with 100 runs has been performed to display the robustness of the filter. Each plot shows the error associated with a quaternion, evaluated as the difference between the true and the estimated value. The quaternions have a large initial uncertainty that is rapidly reduced to steady-state values. The filter is able to track the attitude of the CubeSat, without diverging, as shown by the magenta lines that represent the three standard deviations error boundaries. 
\begin{figure}[!htb]
    \centering
    \includegraphics[width=1.0\linewidth]{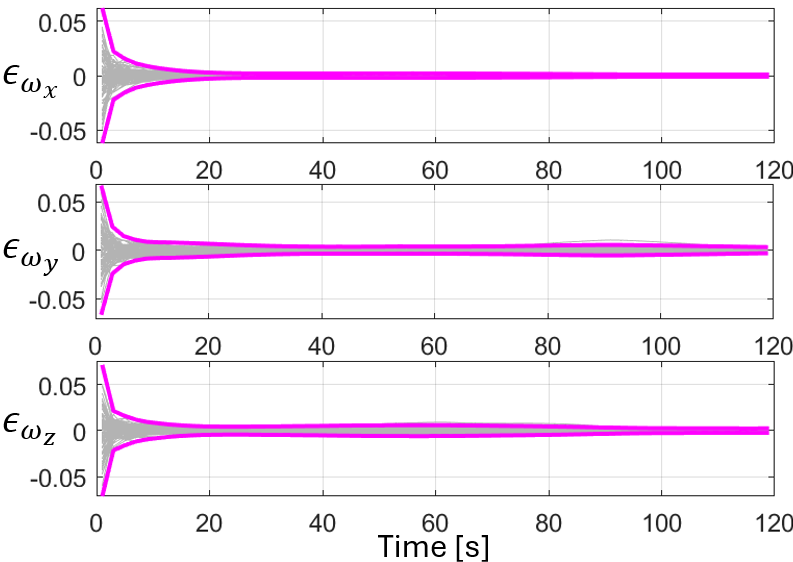}
    \caption{Monte Carlo analysis for the angular velocity error.}
    \label{fig:montew}
\end{figure}
Figure \ref{fig:montew} reports the estimation accuracy for the angular velocity along the whole 120 seconds simulation. Once again, the large initial uncertainty from the initial covariance is rapidly reduced thanks to the gyro measurement and the flow update.  
\begin{figure}[!htb]
    \centering
    \includegraphics[width=1.0\linewidth]{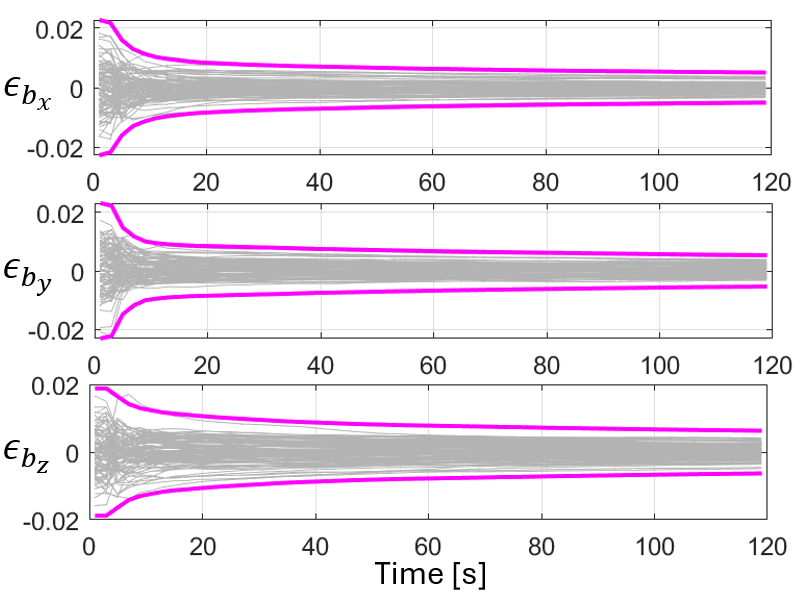}
    \caption{Monte Carlo analysis for the bias estimation error.}
    \label{fig:monteb}
\end{figure}
Lastly, Fig. \ref{fig:monteb} reports the estimation of the gyroscope bias in the three components. The initial bias is estimated correctly, and the error settles below measurement noise levels. Thus, the observability of the bias enhances the performance of the filter, as it provides a more accurate angular velocity estimate and, consequently, attitude.

\begin{figure}[!htb]
    \centering
    \includegraphics[width=1.0\linewidth]{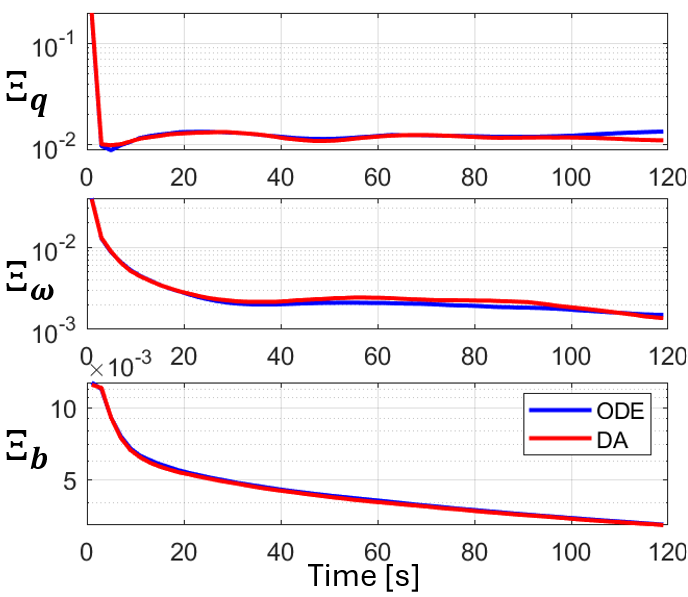}
    \caption{RMSE comparison between ODE and DA solution}
    \label{fig:rmse}
\end{figure}

The performance of the DA flow update is compared to its ODE counterpart from \cite{michaelson2024particle}. Therefore, a Root Mean Square Error (RMSE) analysis from the Monte Carlo results has been carried out. For each $k$th time step, the quaternions RMSE, angular velocity RMSE, and bias RMSE have been evaluated as 
\begin{align}
    \Xi_{\mbf q,k} &= \sqrt{\dfrac{1}{N_{MC}} \sum_{i=1}^{N_{MC}} (\mbf q_{T,k}^{(i)} - \hat{\mbf q}_k^{(i)+})^T(\mbf q_{T,k}^{(i)} - \hat{\mbf q}_k^{(i)+})   }\\
    \Xi_{\boldsymbol{\omega},k} &= \sqrt{\dfrac{1}{N_{MC}} \sum_{i=1}^{N_{MC}} (\boldsymbol{\omega}_{T,k}^{(i)} - \hat{\boldsymbol{\omega}}_k^{(i)+})^T(\boldsymbol{\omega}_{T,k}^{(i)} - \hat{\boldsymbol{\omega}}_k^{(i)+})   }\\
    \Xi_{\mbf b,k} &= \sqrt{\dfrac{1}{N_{MC}} \sum_{i=1}^{N_{MC}} (\mbf b_{T,k}^{(i)} - \hat{\mbf b}_k^{(i)+})^T(\mbf b_{T,k}^{(i)} - \hat{\mbf b}_k^{(i)+})   }
\end{align}
where $N_{MC}$ indicates the number of Monte Carlo simulations. The $\Xi$ parameter is a scalar risk index based on the outer product of the error vector from the estimated state. Figure \ref{fig:rmse} shows that the ODE solution and the DA solution from DARUFF hold extremely similar levels of accuracy in their estimates, as the RMSE lines for the three parts of the state vector overlap for the whole simulation period. Therefore, the DA particle flow filter, implemented with a second-order expansion, provides the same performance as the original ODE derivation.

\begin{figure}[!htb]
    \centering
    \includegraphics[width=1.0\linewidth]{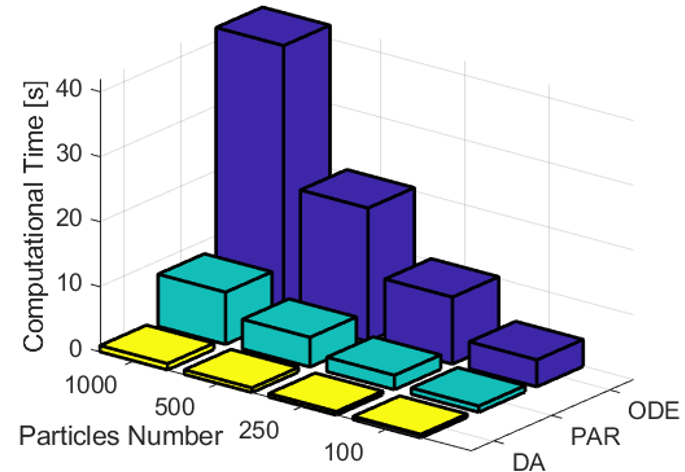}
    \caption{Computational time requested depending on ensemble size}
    \label{fig:timebar}
\end{figure}

The numerical results show how the DA map implementation is a valid substitute for the floating point implementation. However, the major benefit of the novel polynomial technique dwells in the reduced computational burden requested by the system CPU, making it ideally suitable for on-board applications. The computational time comparison has been carried out and reported in Fig. \ref{fig:timebar} as a function of the ensemble size of the particles of the filter. The particles number reported is ``particles per dimension," meaning that the value 100 corresponds to $100n$ total particles. 
The bars comparison includes the nominal numerical integration of the flow, ``ODE", the polynomial map integration and evaluation of flow, ``DA," and an optimized parallelization of the integration of flow and dynamics among the 12 cores of the simulating machine, ``PAR." The DA solution is the fastest among the three particle flow filters, as the yellow bars are well shorter than the ODE and PAR counterparts for each ensemble size. 

\begin{figure}[!htb]
    \centering
    \includegraphics[width=1.0\linewidth]{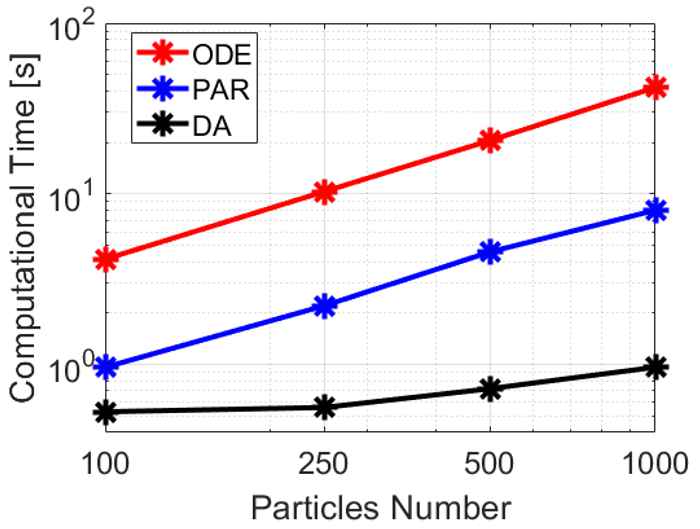}
    \caption{Computational time increase slope comparison}
    \label{fig:timelin}
\end{figure}
A more accurate description is offered by Fig. \ref{fig:timelin}, where the computational time analysis is reported in a double logarithmic scale. This representation highlights two important conclusions: the DA solution is an order of magnitude faster than the optimized parallelized filter, and almost two orders faster than the basic ODE implementation. The second point is the different slope that relates time to the ensemble size. As the filters require a larger number of particles, the DA computational time increases at a slower rate than the parallelized and classical ODE implementation. The more efficient rate is due to the addition of polynomial evaluations rather than numerical integrations. This result implies that the DA solution is particularly efficient for high dimensional applications, where particle filters are usually disregarded due to the computationally heavy trait that makes them unfeasible for low-power hardware. 

\section{Conclusion}\label{sec9}
This paper proposes a different perspective to the continuous recursive measurement update, which becomes a particle flow update. Thanks to the DA representation, the flow is approximated with a series of Taylor polynomials centered at the prior state mean that map deviations to represent the posterior PDF. The result is a computationally efficient flow that substitutes integrations with polynomial evaluations. 

When prediction is included, and a complete filtering algorithm derived, the DA implementation leverages its representation to combine maps as function compositions. That is, the prediction STPM and the flow update map can be merged together in the DA framework to create a single polynomial map that connects each particle from its original location directly to the propagated and updated location, with a single polynomial evaluation.

This new technique has been applied to a toy problem to offer an easy visualization of the particle flow so that the DA implementation reaches the same level of accuracy as the numerical integration counterpart. Afterward, an attitude determination application shows DARUFF as an accurate and robust filter to track the attitude of a CubeSat, keeping a light burden on the CPU due to its reduced complexity. 

In future developments, the diffusion term of the flow will be added to the current derivation that employs just the drift term in the flow differential equation. Moreover, it is possible to use DA to relax the linearization approximation and derive a new set of flow ODE that calculate moments directly in the DA framework.

\bibliographystyle{ieeetr}
\bibliography{references.bib}

\end{document}